\documentclass[runningheads]{llncs}

\usepackage{graphicx}
\usepackage{tikz}
\usepackage{comment}
\usepackage{amsmath,amssymb} 
\usepackage{color}
\usepackage{epsfig}
\usepackage{lineno}
\usepackage{lipsum}

\usepackage[breaklinks=true,bookmarks=false]{hyperref}

\newsavebox\CBox
\def\textBF#1{\sbox\CBox{#1}\resizebox{\wd\CBox}{\ht\CBox}{\textbf{#1}}}

\newcommand{\etal}{{\emph{et al.\ }}}

\begin{document}
\pagestyle{headings}
\mainmatter
\def\ECCVSubNumber{102}  

\title{AIM 2020 Challenge on Rendering Realistic Bokeh } 

\titlerunning{AIM 2020 Challenge on Rendering Realistic Bokeh}
%
%
\author{Andrey Ignatov \and Radu Timofte \and
Ming Qian \and Congyu Qiao \and Jiamin Lin \and Zhenyu Guo \and Chenghua Li \and Cong Leng \and Jian Cheng \and
Juewen Peng \and Xianrui Luo \and Ke Xian \and Zijin Wu \and Zhiguo Cao \and
Densen Puthussery \and Jiji C V \and
Hrishikesh P S \and Melvin Kuriakose \and
Saikat Dutta \and Sourya Dipta Das \and Nisarg A. Shah \and
Kuldeep Purohit \and Praveen Kandula \and Maitreya Suin \and A. N. Rajagopalan \and
Saagara M B \and Minnu A L \and
Sanjana A R \and Praseeda S \and
Ge Wu \and Xueqin Chen \and Tengyao Wang \and
Max Zheng \and Hulk Wong \and Jay Zou $^*$
}

\institute{}
\authorrunning{A. Ignatov,  R. Timofte et al.}

\maketitle

\begin{abstract}
This paper reviews the second AIM realistic bokeh effect rendering challenge and provides the description of the proposed solutions and results. The participating teams were solving a real-world bokeh simulation problem, where the goal was to learn a realistic shallow focus technique using a large-scale EBB! bokeh dataset consisting of 5K shallow / wide depth-of-field image pairs captured using the Canon 7D  DSLR  camera. The participants had to render bokeh effect based on only one single frame without any additional data from other cameras or sensors. The target metric used in this challenge combined the runtime and the perceptual quality of the solutions measured in the user study. To ensure the efficiency of the submitted models, we measured their runtime on standard desktop CPUs as well as were running the models on smartphone GPUs. The proposed solutions significantly improved the baseline results, defining the state-of-the-art for practical bokeh effect rendering problem.
\end{abstract}

\section{Introduction}
\label{sec:introduction}

\let\thefootnote\relax\footnotetext{
$^*$ A. Ignatov and R. Timofte (\{andrey,radu.timofte\}@vision.ee.ethz.ch, ETH Zurich) are the challenge organizers, while the other authors participated in the challenge.\\
The Appendix~\ref{sec:affiliations} contains the authors' teams and affiliations.\\
AIM 2020 webpage: \url{https://data.vision.ee.ethz.ch/cvl/aim20/}}

The advances in image manipulation tasks are impressive. In particular, the image manipulation related to portable devices such as smartphone cameras has recently faced an interest boost from the research community to match the users' demands. Multiple novel solutions were proposed in the literature for various tasks, such as image quality enhancement~\cite{ignatov2017dslr,ignatov2018pirm,Timofte_2018_CVPR_Workshops,NTIRE_Dehazing_2019}, style transfer~\cite{gatys2016image,johnson2016perceptual,luan2017deep},
learning of an image signal processor (ISP)~\cite{ignatov2020replacing}, photo segmentation and blurring~\cite{wadhwa2018synthetic,shen2016automatic,chen2017deeplab,badrinarayanan2017segnet}, etc. Moreover, modern mobile devices got powerful GPUs and NPUs that are well suitable for running the proposed deep learning models~\cite{ignatov2018ai,ignatov2019ai}.

Rendering an automatic bokeh effect has been one of the most popular topics over past few years, with many solutions that are now included within the majority of smartphone camera applications.
In 2014, a seminal work on portrait segmentation~\cite{GoogleBlur2014} was published, and substantial improvements in segmentation accuracy were reported in many subsequent papers~\cite{shen2016automatic,xu2017deep}.
Wadhwa~\etal~\cite{wadhwa2018synthetic} provided a detailed description of the synthetic depth-of-field rendering method found in the Google Pixel phones and inspired further development in this field.

The AIM 2020 challenge on rendering realistic bokeh builds upon the success of the previous AIM 2019 challenge~\cite{ignatov2019aim}, and advances the benchmarking of example-based single image bokeh effect rendering by introducing two tracks with evaluation on several recent-generation desktop CPUs and smartphone GPUs. The AIM 2020 challenge uses the large-scale EBB!~\cite{ignatov2020rendering} dataset consisting of photo pairs with shallow and wide depth-of-field captured using the Canon 70D DSLR camera. Quantitative and qualitative visual results as well as the inference time and efficiency are used for ranking the proposed solutions.
The challenge, the corresponding dataset, the results and the proposed methods are described and discussed in the next sections.

This challenge is one of the AIM 2020 associated challenges on:
scene relighting and illumination estimation~\cite{elhelou2020aim_relighting}, image extreme inpainting~\cite{ntavelis2020aim_inpainting}, learned image signal processing pipeline~\cite{ignatov2020aim_ISP}, rendering realistic bokeh~\cite{ignatov2020aim_bokeh}, real image super-resolution~\cite{wei2020aim_realSR}, efficient super-resolution~\cite{zhang2020aim_efficientSR}, video temporal super-resolution~\cite{son2020aim_VTSR} and video extreme super-resolution~\cite{fuoli2020aim_VXSR}.

\section{AIM 2020 Challenge on  Realistic Bokeh}

The objectives of the AIM 2020 challenge on rendering realistic bokeh effect is to promote realistic settings as defined by the \textit{EBB!} Bokeh dataset, to push the state-of-the-art in synthetic shallow depth-of-field rendering, and to ensure that the final solutions are efficient enough to run both on desktop and mobile hardware.

\subsection{\textit{Everything is Better with Bokeh!} Dataset}

\begin{figure*}[t!]
\centering
\includegraphics[width=1.0\linewidth]{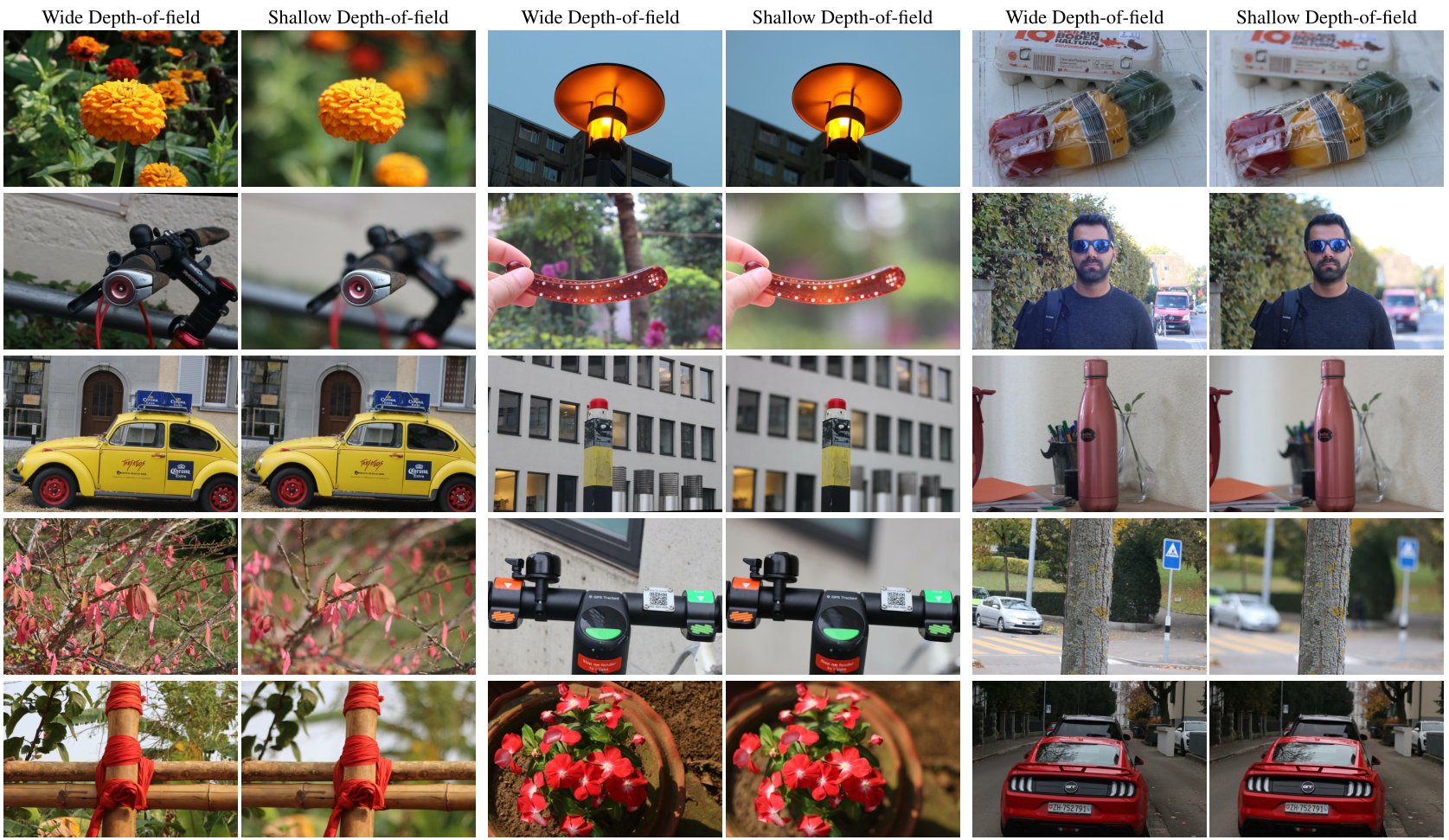}
\caption{Sample wide and shallow depth-of-field image pairs from the EBB! dataset.}
\label{fig:sample_images}
\end{figure*}

One of the biggest challenges in the bokeh rendering task is to get high-quality real data that can be used for training deep models. To tackle this problem, we used a large-scale \textit{Everything is Better with Bokeh!} (EBB!) dataset presented in~\cite{ignatov2020rendering} that is containing more than 10 thousand images collected in the wild during several months. By controlling the aperture size of the lens, images with shallow and wide depth-of-field were taken. In each photo pair, the first image was captured with a narrow aperture (f/16) that results in a normal sharp photo, whereas the second one was shot using the highest aperture (f/1.8) leading to a strong bokeh effect. The photos were taken during the daytime in a wide variety of places and in various illumination and weather conditions. The photos were captured in automatic mode, the default settings were used throughout the entire collection procedure. An example set of collected images is presented in Figure~\ref{fig:sample_images}.

The captured image pairs are not aligned exactly, therefore they were first matched using SIFT keypoints and RANSAC method same as in~\cite{ignatov2017dslr}. The resulting images were then cropped to their intersection part and downscaled so that their final height is equal to 1024 pixels. From the resulting 10 thousand images, 200 image pairs were reserved for testing, while the other 4.8 thousand photo pairs can be used for training and validation.

\subsection{Tracks and Competitions}

The challenge consists of the following phases:

\vspace{-0.8mm}
\begin{enumerate}
\setlength\itemsep{-0.2mm}
\item[i] \textit{development:} the participants get access to the data;
\item[ii] \textit{validation:} the participants have the opportunity to validate their solutions on the server and compare the results on the validation leaderboard;
\item[iii] \textit{test:} the participants submit their final results, models, and factsheets.
\end{enumerate}
\vspace{-0.8mm}

All submitted solutions were evaluated based on the following measures:

\vspace{-0.8mm}
\begin{itemize}
\setlength\itemsep{-0.2mm}
\item PSNR measuring fidelity score,
\item SSIM, a proxy for perceptual score,
\item The runtime of the submitted models on desktop CPUs and mobile GPUs,
\item MOS scores measured in the user study for explicit image quality assessment.
\end{itemize}
\vspace{-0.8mm}

The AIM 2020 challenge on  realistic bokeh consists of two tracks. In the first ``CPU'' track, the target was to produce a model which runtime is optimized for standard desktop CPUs. In the second, ``Smartphone GPU'' track, the goal was to develop a TensorFlow Lite~\cite{TensorFlowLite2018} compatible solution that was tested on several mobile GPUs using a publicly available$^1$\footnotetext{$^1$ \url{http://ai-benchmark.com}} \textit{AI Benchmark} application~\cite{ignatov2019ai} and an OpenCL-based TFLite GPU delegate~\cite{TFLite2019GPU}. During the development and validation phases, the quantitative performance of the solutions was measured by PSNR and SSIM metric.  Since SSIM and PSNR scores are not reflecting many aspects of real quality of the resulted images, during the final test phase we evaluated the solutions based on their Mean Opinion Scores (MOS). For this, we conducted a user study evaluating the visual results of all proposed methods. The users were asked to rate the quality of each submitted solution by selecting one of the five quality levels (5 - comparable perceptual quality, 4 - slightly worse, 3 - notably worse, 2 - poor perceptual quality, 1 - completely corrupted image) for each method result in comparison with the original Canon images exhibiting bokeh effect. The expressed preferences were averaged per each test image and then per each method to obtain the final MOS.

\section{Challenge Results}

\begin{table*}[tbh!]
\centering
\resizebox{\linewidth}{!}
{
\begin{tabular}{l|c|cc|cccc}
\multicolumn{2}{c|}{}&\multicolumn{2}{c|}{Factsheet Info}&\multicolumn{4}{c}{Track 1: Desktop CPU}\\

\hline
Team \, & \, Author \, & \, Framework \, & \, Training Hardware, GPU \, & \, Avg. Runtime, s \, & \, PSNR$\uparrow$ \, & \, SSIM$\uparrow$ \, & \, MOS$\uparrow$ \\
\hline
\hline
Airia-bokeh & \, MingQian \, & TensorFlow & Nvidia TITAN RTX & 5.52 & 23.58 & 0.8770 & \textBF{4.2} \\
AIA-Smart & \, JuewenPeng \, & PyTorch & GeForce GTX 1080 & 1.71 & 23.56 & 0.8829 & 3.8 \\
CET\_SP & \, memelvin99 \, & TensorFlow & Nvidia Tesla P100 & 1.17 & 21.91 & 0.8201 & 3.3 \\
CET\_CVLab & \, Densen \, & TensorFlow & Nvidia Tesla P100 & 1.17 & 23.05 & 0.8591 & 3.2 \\
Team Horizon & \, tensorcat \, & PyTorch & GeForce GTX 1080 Ti & 19.27 & 23.27 & 0.8818 & 3.2 \\
IPCV\_IITM & \, ms\_ipcv \, & PyTorch & NVIDIA Titan X & 27.24 & \textBF{23.77} & \textBF{0.8866} & 2.5 \\
CET21\_CV & \, SaagaraMB \, & TensorFlow & Nvidia Tesla P100 & \textBF{0.74} & 22.80 & 0.8628 & 1.3 \\
CET\_ECE & \, Sanjana.A.R \, & TensorFlow & Nvidia Tesla P100 & \textBF{0.74} & 22.85 & 0.8629 & 1.2 \\
xuehuapiaopiao-team \, & \, xuehuapiaopiao \, & TensorFlow & GeForce GTX 1080 Ti & - & 22.98 & 0.8758 & - * \\
Terminator & \, Max\_zheng \, & TensorFlow & GeForce GTX 1080 Ti & - & 23.04 & 0.8756 & - * \\
\end{tabular}
}
\vspace{2.6mm}
\caption{\small{AIM 2020 realistic bokeh rendering challenge, CPU Track: results and final rankings. The results are sorted based on the MOS scores. $^*$ - These teams submitted solutions that are using pre-computed depth maps and therefore were excluded from the final evaluation phase.}}
\label{tab:results}
\end{table*}

\begin{table*}[tbh!]
\centering
\resizebox{\linewidth}{!}
{
\begin{tabular}{l|c|cc|cccc}
\multicolumn{2}{c|}{}&\multicolumn{2}{c|}{Factsheet Info}&\multicolumn{4}{c}{Track 2: Smartphone GPU}\\

\hline
Team \, & \, Author \, & \, Framework \, & \, Training Hardware, GPU \, & \, Avg. Runtime, s \, & \, PSNR$\uparrow$ \, & \, SSIM$\uparrow$ \, & \, MOS$\uparrow$ \\
\hline
\hline
Airia-bokeh & \, MingQian \, & TensorFlow & Nvidia TITAN RTX & \textBF{1.52} & \textBF{23.58} & 0.8770 & \textBF{4.2} \\
AIA-Smart & \, JuewenPeng \, & PyTorch & GeForce GTX 1080 & 15.2 & 22.94 & \textBF{0.8842} & 4.0 \\
CET\_CVLab & \, Densen \, & TensorFlow & Nvidia Tesla P100 & 2.75 & 23.05 & 0.8591 & 3.2 \\
Team Horizon \, & \, tensorcat \, & PyTorch & GeForce GTX 1080 Ti & - * & 23.27 & 0.8818 & 3.2 \\
\end{tabular}
}
\vspace{2.6mm}
\caption{\small{AIM 2020 realistic bokeh rendering challenge, GPU Track: results and final rankings. The results are sorted based on the MOS scores. The model submitted by the Team Horizon was unable to run on mobile GPUs due to NCHW channel order that is currently not supported by the TensorFlow Lite GPU delegate.}}
\label{tab:results_gpu}
\end{table*}

The Track 1 of the challenge attracted more than 110 registered participants and the Track 2 more than 80. However, only 9 teams provided results in the final phase together with factsheets and codes for reproducibility.
Tables~\ref{tab:results} and~\ref{tab:results_gpu} summarize the final challenge results in terms of PSNR, SSIM and MOS scores for each submitted solution in addition to self-reported hardware / software configurations and runtimes. Short descriptions of the proposed solutions are provided in section~\ref{sec:solutions}, and the team details (contact email, members and affiliations) are listed in Appendix~\ref{sec:affiliations}.


\subsection{Architectures and Main Ideas}

All the proposed methods are relying on end-to-end deep learning-based solutions. Almost all submitted models have a multi-scale encoder-decoder architecture and are processing the images at several scales. This allows to achieve a significantly faster runtime as all heavy image processing is done on images of low resolution, as well as adds the possibility of introducing heavy global image manipulations. The majority of teams were using the $L1$, SSIM / MS-SSIM, VGG-based, Sobel and Charbonnier loss functions, while team Airia-bokeh demonstrated that a proper adversarial loss can significantly boost the quality of the resulting bokeh effect. Almost all teams were using the Adam optimizer~\cite{kingma2014adam} to train deep learning models and TensorFlow or PyTorch frameworks to implement and train the networks.

\subsection{Performance}

\paragraph{Quality.}
Airia-bokeh is the winner of the AIM 2020 challenge on rendering realistic bokeh. Airia-bokeh ranks the best in perceptual quality in both track 1 and track 2 with the same solution (deep model).
Only one team -- AIA-Smart, -- submitted different models / solutions for the two tracks of the challenge. Surprisingly, the solution submitted for evaluation on smartphone GPU (Track 2) obtained better SSIM and MOS results than the one for CPU (Track 1). They are coming second to Airia-bokeh in the MOS score while reporting the best SSIM (0.8842) in Track 2. As expected, the perceptual ranking according to the MOS does not strongly correlate with the fidelity measures such as PSNR and SSIM. In particular, IPCV\_IITM team ranks first in terms of SSIM and PSNR but only sixth in terms of perceptual quality (Track 1). Interestingly, the CET\_SP team has the lowest fidelity (PSNR) and SSIM results, though comes third in perceptual quality (MOS).

\paragraph{Runtime.} The measured average runtimes of the proposed solutions on standard Nvidia GPU cards (CPU Track 1) vary from ~0.7s to more than 27s per single image. The fastest solutions ($\sim$0.7s) are also among the worst performing in perceptual ranking, while the top fidelity method proposed by IPCV\_IITM requires 27s, and the top perceptual methods of Airia-bokeh and of AIA-Smart require 5.52s and 1.71s, respectively.
When it comes to the solutions proposed in Track 2 evaluated on smartphone GPUs, the best perceptual quality solution of Airia-bokeh is also with the lowest inference time, 1.52s. We conclude that the proposed solutions do not meet the requirements for real-time applications on the current generation of smartphones, thus all processing should be done in the background after the image is obtained / captured.

\subsection{Discussion}

With the AIM 2020 challenge, we went further compared to the previously held challenges and aimed at solutions meant to run efficiently on desktop and smartphone hardware.
The challenge employed the EBB!~\cite{ignatov2020rendering}, a novel large dataset containing paired and aligned low- and high-aperture photos captured with a high-end Canon 70D DSLR camera. Several of the proposed approaches produced results with good perceptual quality and runtime suitable for on-device image processing. These methods are gauging the state-of-the-art for the practical bokeh synthesis task learned from pairs of real exemplars.

\section{Challenge Methods and Teams}
\label{sec:solutions}

This section describes solutions submitted by all teams participating in the final stage of the AIM 2020 realistic bokeh rendering challenge.

\smallskip

\subsection{Airia-bokeh}

\begin{figure}[h!]
\centering
\resizebox{1.0\linewidth}{!}
{
\includegraphics[width=1.0\linewidth]{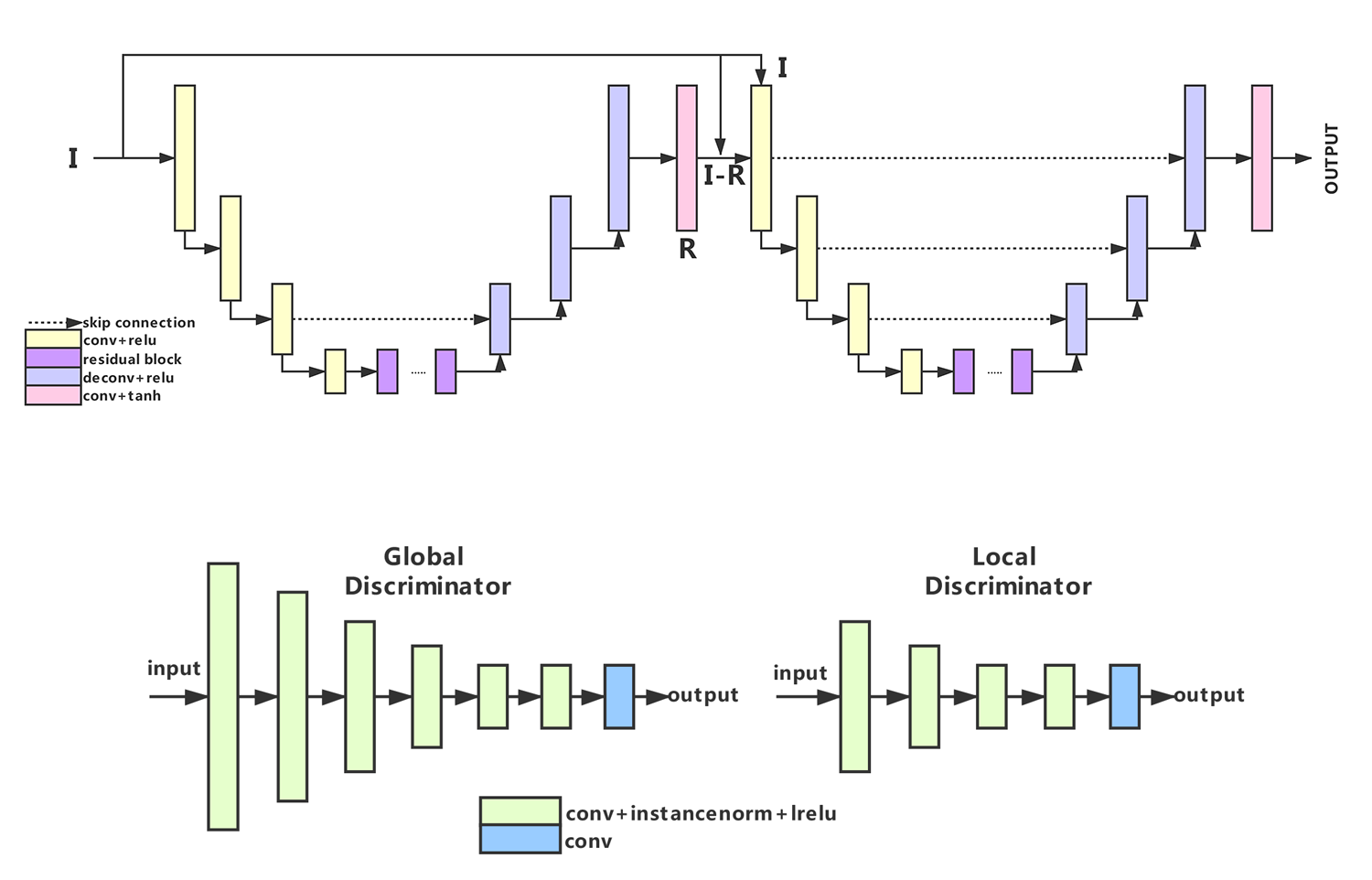}
}
\caption{\small{Bokeh-Glass Network network (top) and PatchGAN-based discriminators (bottom) proposed by Airia-bokeh team.}}
\label{fig:Airia}
\end{figure}

Team Airia-bokeh proposed a Bokeh-Glass Network (BG-Net)~\cite{qian2020bggan} model for rendering realistic bokeh that is illustrated in Fig.~\ref{fig:Airia}. The model consists of two stacked U-Net based networks that were first trained separately using a combination of the $L_1$ and SSIM losses (with weights 0.5 and 1, respectively). During the second stage, two PatchGAN~\cite{zhu2017unpaired} discriminators with different receptive fields were added to improve the quality of the produced images. The generator and the discriminator were trained together using the WGAN-GP algorithm with a batch size of 1. The authors have additionally enhanced the EBB! dataset by removing some image pairs that did not correspond in color or were not in focus.

\subsection{AIA-Smart}

\begin{figure}[h!]
\centering
\resizebox{1.0\linewidth}{!}
{
\includegraphics[width=1.0\linewidth]{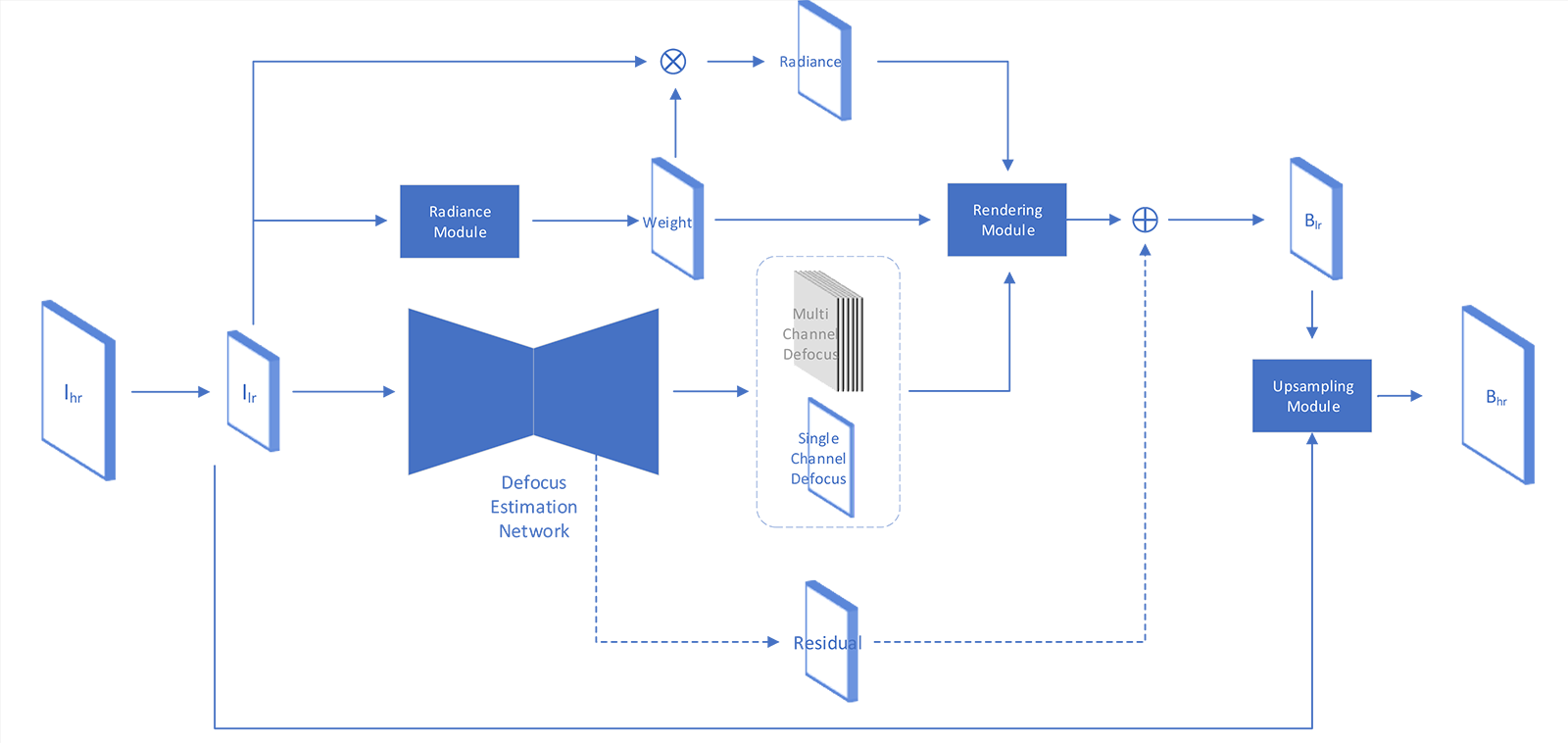}
}
\caption{\small{AIA-Smart network consisting of the defocus estimation, radiance, rendering and upsampling modules.}}
\label{fig:AIA}
\end{figure}

The solution of AIA-Smart team is based on defocus map estimation~\cite{luo2020aim_defocus}. The proposed architecture consists of 4 modules (Fig.~\ref{fig:AIA}): defocus estimation, radiance, rendering and upsampling modules. Defocus estimation module is used to predict a defocus map, which works as a guidance for defocus rendering. Radiance module calculates the weight map used for estimating the weight of each pixel in the rendering process. In the rendering module, the low resolution bokeh result can be obtained with the input of the radiance map, weight map and defocus map using the refocusing pipeline proposed in~\cite{busam2019sterefo}. In the upsampling module,
the low resolution bokeh result and the high resolution original image are combined to generate the final full-resolution bokeh image.

The training of the network can be divided into 4 stages: 1) predicting the layered defocus maps and render the bokeh result on 1/4 of the original resolution, 2) rendering bokeh effect at 1/2 resolution while using the pretrained network from the first stage to refine the details around foreground boundaries, 3) replacing the multi-channel classification layer with a single-channel regression layer in the defocus estimation module to generate the pleasing 'Circle of Confusion', and 4) rendering the image at 1/2 resolution, upsampling the result by bilinear interpolation and calculating a soft foreground mask from the predicted single-channel defocus map. Finally, the foreground objects of the original image are covered on the rendering result to make the foreground more clear.

During the first stage, the model is trained with a combination of the $L_1$, perceptual, SSIM and gradient loss functions using the images of resolution 256$\times$256 pixels. The initial learning rate is set to $1e-4$ with a decay-cycle of 30 epochs. At the second and the third stages, the model is fine-tuned on 512$\times$512 pixel images using the same set of loss functions.

\subsection{CET\_CVLab and CET\_SP}

\begin{figure}[h!]
\centering
\resizebox{1.0\linewidth}{!}
{
\includegraphics[width=1.0\linewidth]{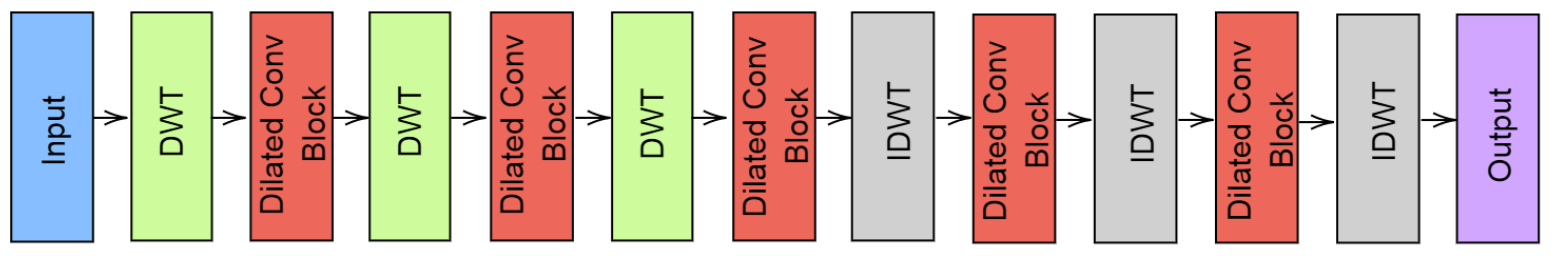}
}
\caption{\small{Dilated Wavelet CNN model used by CET\_CVLab and CET\_SP teams.}}
\label{fig:CETCVLab}
\end{figure}

Both CET\_CVLab and CET\_SP teams used the same U-Net based Dilated Wavelet CNN model (Fig.~\ref{fig:CETCVLab}) for generating bokeh images. In this network, the standard downsampling and upsampling operations are replaced with a discrete wavelet transform based (DWT)  decomposition to minimize the information loss in these layers. The proposed methodology is computationally efficient and is based on the multi-level wavelet-CNN (MWCNN) proposed in~\cite{liu2018multi}.

CET\_SP trained the model with a combination of the Charbonnier and perceptual VGG loss, while CET\_CVLab additionally used Sobel and Grayscale ($L_1$ distance between the grayscale images) loss functions. Both models were optimized using the Adam algorithm with a batch size of 10 for 600 and 500 epoch, respectively.

\subsection{Team Horizon}

\begin{figure}[h!]
\centering
\resizebox{1.0\linewidth}{!}
{
\includegraphics[width=1.0\linewidth]{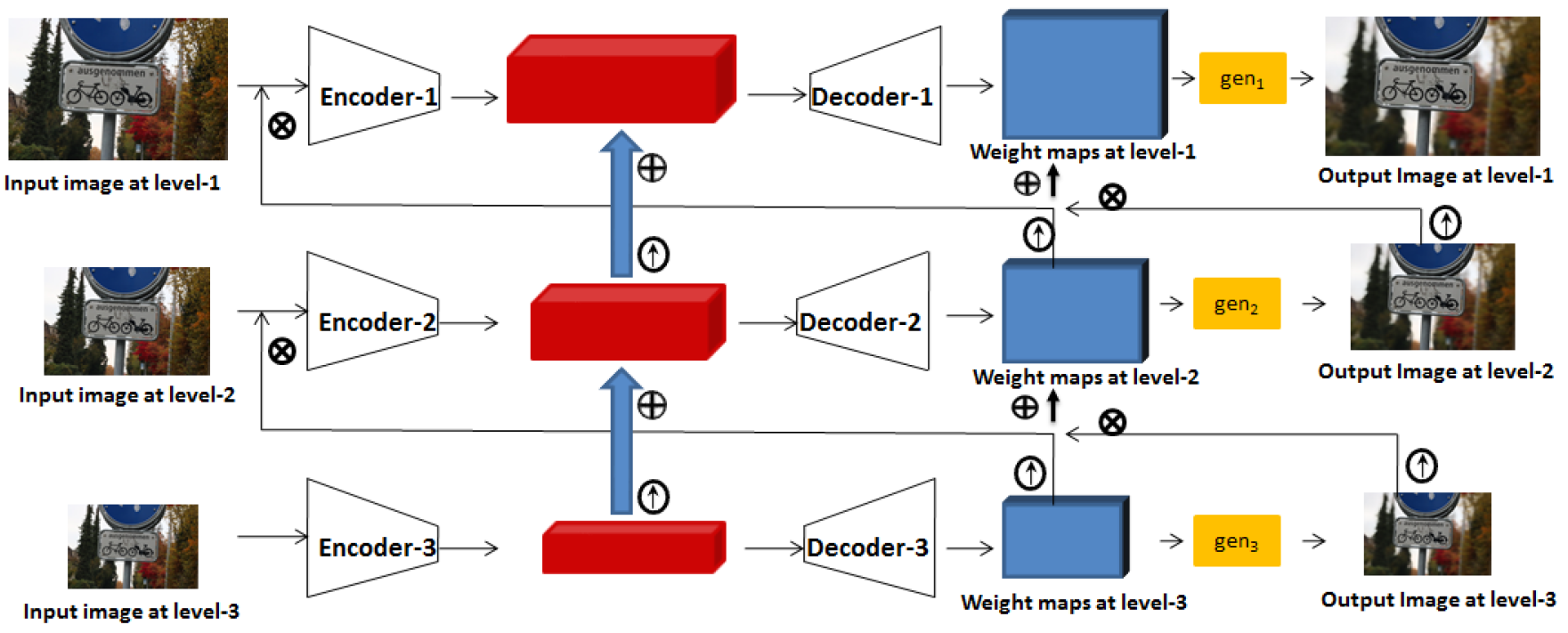}
}
\caption{\small{A multiscale encoder-decoder based model proposed by Team Horizon.}}
\label{fig:Horizon}
\end{figure}

The authors proposed an encoder-decoder based model shown in Fig.~\ref{fig:Horizon} that is trained at several scales. At each level, the encoder-decoder module is producing the weight maps that are used together with the input image by the bokeh generation module to render the bokeh image. Generated weight maps and bokeh images are then upscaled and concatenated with the input image in the next level, while the upscaled encoded features are added to the corresponding encoded features used in the next level. The model is trained with a combination of the MS-SSIM and SSIM loss functions.

\subsection{IPCV\_IITM}

\begin{figure}[h!]
\centering
\resizebox{1.0\linewidth}{!}
{
\includegraphics[width=1.0\linewidth]{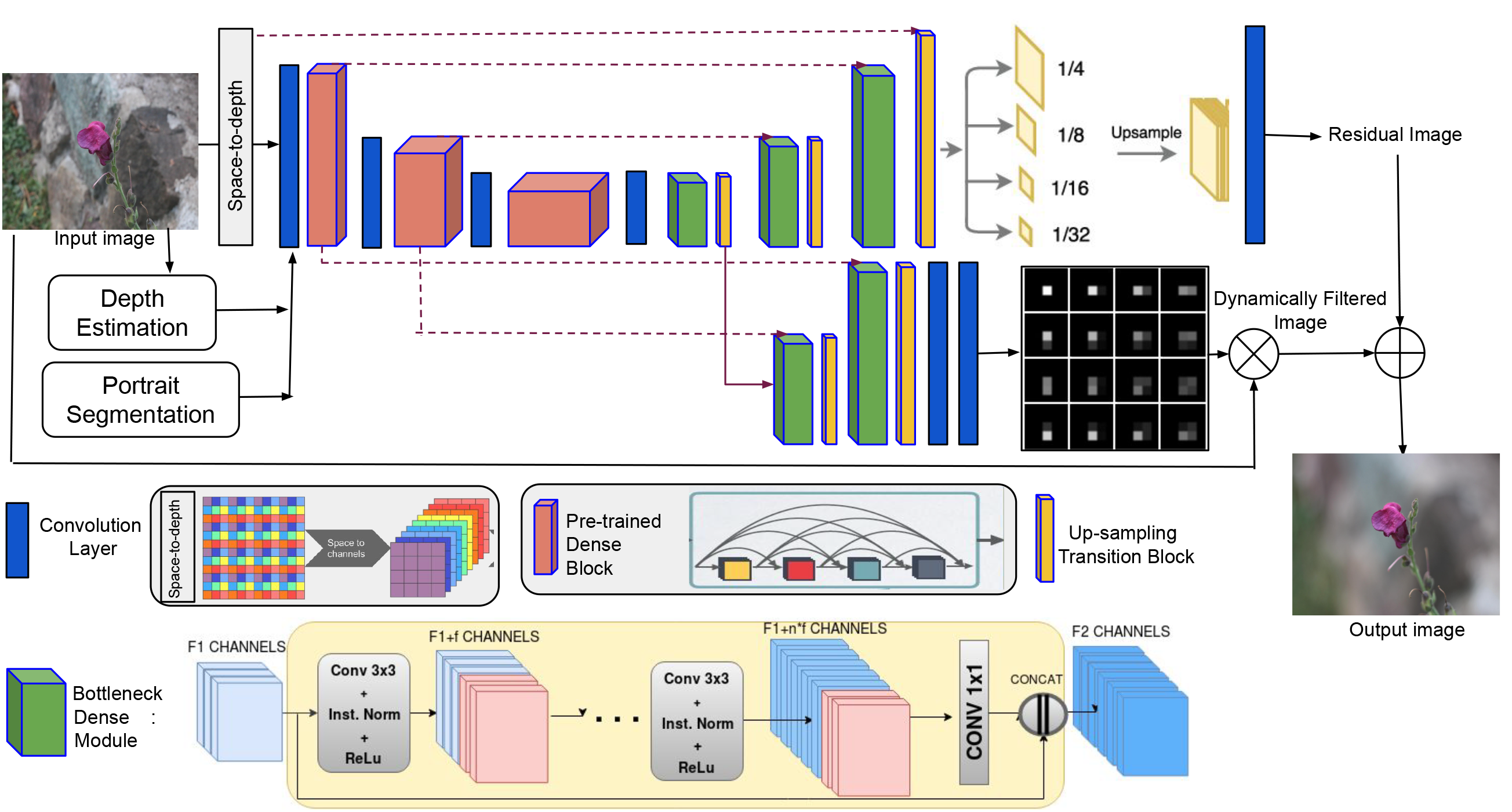}
}
\caption{\small{Depth-guided Dynamic Filtering Dense Network proposed by IPCV\_IITM.}}
\label{fig:IPCV}
\end{figure}

The authors proposed a depth-guided dynamic filtering dense network for rendering shallow depth-of-field (Fig.~\ref{fig:IPCV}). At the onset, the network uses a space-to-depth module that divides each input channel into a number of blocks concatenated along the channel dimension. The output of this layer is concatenated with the outputs of the pre-trained depth estimation~\cite{li2018megadepth} and salient object segmentation~\cite{hou2017deeply} networks to achieve more accurate rendering results. The resulting feature maps are passed to a U-net~\cite{ronneberger2015u} based encoder consisting of densely connected modules. The first dense-block contains 12 densely-connected layers, the second block~-- 16, and the third one~-- 24 densely-connected layers. The weights of each block are initialized using the DenseNet-121 network~\cite{huang2017densely} trained on the ImageNet dataset. The decoder has two difference branches which outputs are summed to produce the final result. The first branch has a U-net architecture with skip-connections and also consists of densely connected blocks. Its output is enhanced through multi-scale context aggregation through pooling and upsampling at 4 scales. The second branch uses the idea of dynamic filtering~\cite{jia2016dynamic} and generates dynamic blurring filters conditioned on the encoded feature map. These filters are produced locally and on-the-fly depending on the input, the parameters of the filter-generating network are updated during the training.
We refer to~\cite{purohit2019depth} for more details.

\subsection{CET21\_CV}

\begin{figure}[h!]
\centering
\resizebox{1.0\linewidth}{!}
{
\includegraphics[width=1.0\linewidth]{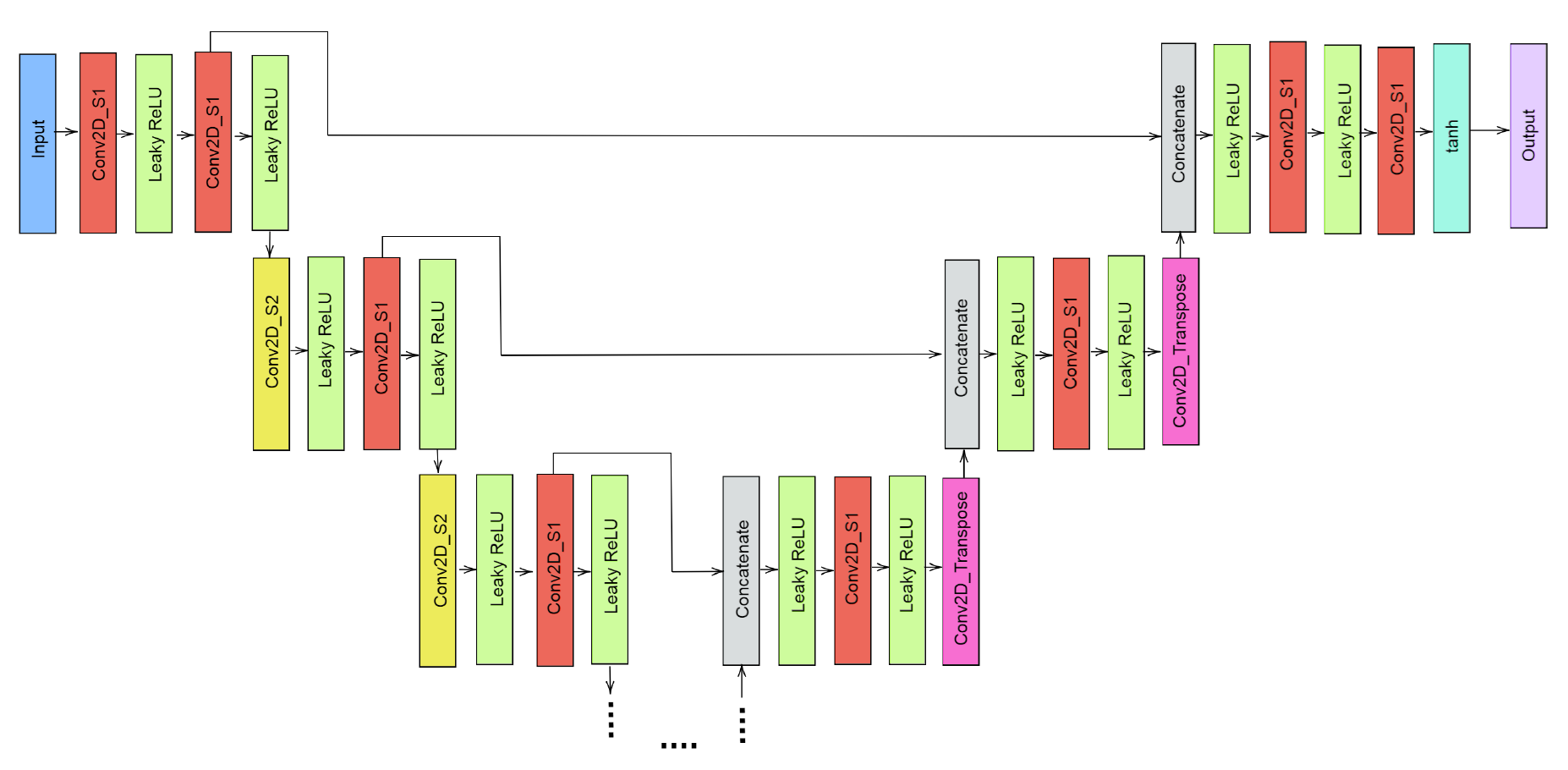}
}
\caption{\small{A modified U-Net model used by CET21\_CV team.}}
\label{fig:CET21}
\end{figure}

CET21\_CV proposed a modified U-Net model depicted in Fig.~\ref{fig:CET21}. Compared to the original U-Net implementation, the authors replaced the max-pooling downsampling operation with a strided convolution layer, and the feature maps from shortcut connections are concatenated before applying the activation functions in the decoder module. \textit{Leaky ReLU} activations are used in the convolutional layers, and the entire model is trained to minimize the mean absolute error loss using the Adam algorithm.

\subsection{CET\_ECE}

\begin{figure}[h!]
\centering
\resizebox{1.0\linewidth}{!}
{
\includegraphics[width=1.0\linewidth]{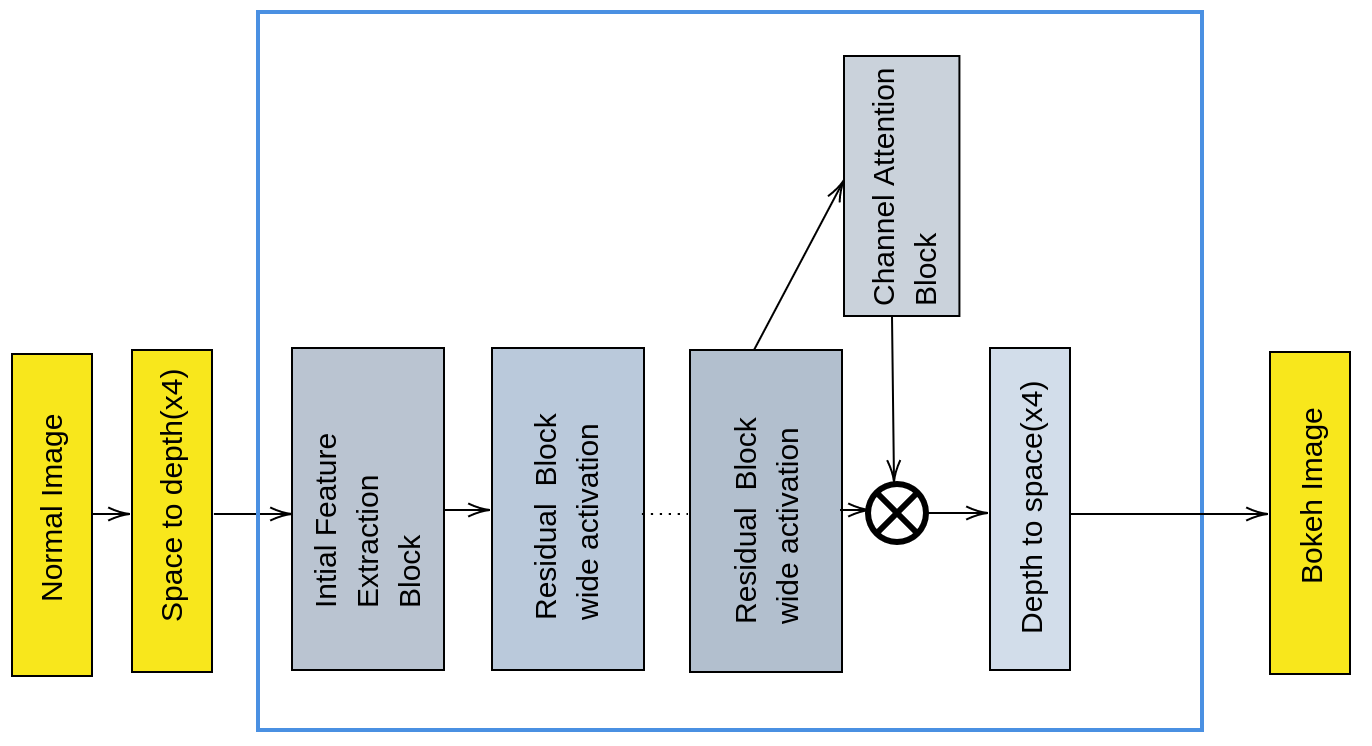}
}
\caption{\small{CET\_ECE's network architecture.}}
\label{fig:ECE}
\end{figure}

The model architecture proposed by CET\_ECE team was inspired from the wide activation~\cite{yu2018wide} and channel attention~\cite{zhang2018image} based networks. The proposed network (Fig.~\ref{fig:ECE}) is generally consisting of 2 block types: a feature extraction block and a series of wide activation residual blocks. To reduce the model complexity and information loss, a space-to-depth layer with a scale factor of 4 is used before the initial feature extraction block, and a depth-to-space operation is used as the last layer of the network. The Charbonnier loss function is used for training the network as it better captures the edge information compared to the MSE loss.

\subsection{Xuehuapiaopiao-team and Terminator}

Both teams used a slightly modified PyNET~\cite{ignatov2020rendering} model for generating bokeh images. While the visual results of the proposed solutions were looking fine, they were relying on depth estimation modules that were not included in the submissions, and therefore were not ranked in the final phase of the challenge.

\section*{Acknowledgments}
We thank the AIM 2020 sponsors: Huawei, MediaTek, Qualcomm, NVIDIA, Google and Computer Vision Lab / ETH Z\"urich.

\newpage

\appendix
\section{Appendix 1: Teams and affiliations}
\label{sec:affiliations}

\bigskip

\noindent{\textbf{AIM 2020 Realistic Bokeh Rendering Challenge Team}}
\small
\bigskip

\noindent
\textBF{Title:} AIM 2020 Challenge on Rendering Realistic Bokeh

\smallskip

\noindent
\textBF{Members:} \hspace{0.7mm} Andrey Ignatov \,--\, \footnotesize{andrey@vision.ee.ethz.ch}, \small

\hspace{10.5mm} Radu Timofte \,--\, \footnotesize{radu.timofte@vision.ee.ethz.ch}

\smallskip

\noindent
\textBF{Affiliations:} \hspace{0.7mm} Computer Vision Lab, ETH Zurich, Switzerland


\bigskip

\smallskip

\noindent
\textbf{Airia-bokeh}
\small
\bigskip

\noindent
\textBF{Title:} BGNet: Bokeh-Glass Network for Rendering Realistic Bokeh

\smallskip

\noindent
\textBF{Members:} \hspace{0.7mm} Ming Qian \,--\, \footnotesize{20181223053@nuist.edu.cn}, \small

\hspace{10.5mm} Congyu Qiao, Jiamin Lin, Zhenyu Guo, Chenghua Li, 

\hspace{10.5mm} Cong Leng, Jian Cheng

\smallskip

\noindent
\textBF{Affiliations:} \hspace{0.7mm} Nanjing Artificial Intelligence Chip Research, Institute of Automation

\hspace{12.9mm} Chinese Academy of Sciences (AiRiA), MAICRO, China

\bigskip

\smallskip

\noindent
\textbf{AIA-Smart}
\small
\bigskip

\noindent
\textBF{Title:} Bokeh Rendering from Defocus Estimation~\cite{luo2020aim_defocus}

\smallskip

\noindent
\textBF{Members:} \hspace{0.7mm} Juewen Peng \,--\, \footnotesize{im.pengjw@gmail.com}, \small

\hspace{10.5mm} Xianrui Luo, Ke Xian, Zijin Wu, Zhiguo Cao

\smallskip

\noindent
\textBF{Affiliations:} \hspace{0.7mm} Huazhong University of Science and Technology, China

\bigskip

\smallskip

\noindent
\textbf{CET\_CVLab}
\small
\bigskip

\noindent
\textBF{Title:} Photorealistic Bokeh Effect Rendering with Dilated Wavelet CNN

\smallskip

\noindent
\textBF{Members:} \hspace{0.7mm} Densen Puthussery \,--\, \footnotesize{puthusserydensen@gmail.com}, \small

\hspace{10.5mm} Jiji C V

\smallskip

\noindent
\textBF{Affiliations:} \hspace{0.7mm} College of Engineering Trivandrum, India

\bigskip

\smallskip

\noindent
\textbf{CET\_SP}
\small
\bigskip

\noindent
\textBF{Title:} Bokeh Effect using VGG based Wavelet CNN

\smallskip

\noindent
\textBF{Members:} \hspace{0.7mm} Hrishikesh P S \,--\, \footnotesize{hrishikeshps@cet.ac.in}, \small

\hspace{10.5mm} Melvin Kuriakose

\smallskip

\noindent
\textBF{Affiliations:} \hspace{0.7mm} College of Engineering Trivandrum, India

\bigskip

\smallskip

\noindent
\textbf{Team Horizon}
\small
\bigskip

\noindent
\textBF{Title:} Deep Multi-scale Hierarchical Network for Bokeh Effect Rendering

\smallskip

\noindent
\textBF{Members:} \hspace{0.7mm} Saikat Dutta \,--\, \footnotesize{cs18s016@smail.iitm.ac.in}, \small

\hspace{10.5mm} Sourya Dipta Das, Nisarg A. Shah

\smallskip

\noindent
\textBF{Affiliations:} \hspace{0.7mm} Indian Institute of Technology Madras, India

\hspace{12.9mm} Jadavpur University,  India

\hspace{12.9mm} Indian Institute of Technology Jodhpur, India

\bigskip

\smallskip

\noindent
\textbf{IPCV\_IITM}
\small
\bigskip

\noindent
\textBF{Title:} Dense Dynamic Filtering Network for Rendering Synthetic Depth-of-Field Effect

\smallskip

\noindent
\textBF{Members:} \hspace{0.7mm} Kuldeep Purohit \,--\, \footnotesize{kuldeeppurohit3@gmail.com}, \small

\hspace{10.5mm} Praveen Kandula, Maitreya Suin, A. N. Rajagopalan

\smallskip

\noindent
\textBF{Affiliations:} \hspace{0.7mm} Indian Institute of Technology Madras, India

\bigskip

\smallskip

\noindent
\textbf{CET21\_CV}
\small
\bigskip

\noindent
\textBF{Title:} Synthetic Bokeh Effect with Modified UNet

\smallskip

\noindent
\textBF{Members:} \hspace{0.7mm} Saagara M B \,--\, \footnotesize{saagara@cet.ac.in}, \small

\hspace{10.5mm} Minnu A L

\smallskip

\noindent
\textBF{Affiliations:} \hspace{0.7mm} College of Engineering Trivandrum, India

\bigskip

\smallskip

\noindent
\textbf{CET\_ECE}
\small
\bigskip

\noindent
\textBF{Title:} Bokeh Effect Rendering with Deep Convolutional Neural Network

\smallskip

\noindent
\textBF{Members:} \hspace{0.7mm} Sanjana A R \,--\, \footnotesize{ar.sanjanaar@gmail.com}, \small

\hspace{10.5mm} Praseeda S

\smallskip

\noindent
\textBF{Affiliations:} \hspace{0.7mm} College of Engineering Trivandrum, India

\bigskip

\smallskip

\noindent
\textbf{Xuehuapiaopiao-team}
\small
\bigskip

\noindent
\textBF{Title:} Multi-scale Bokeh Rendering Network

\smallskip

\noindent
\textBF{Members:} \hspace{0.7mm} Ge Wu \,--\, \footnotesize{1047670389@qq.com}, \small

\hspace{10.5mm} Xueqin Chen, Tengyao Wang

\smallskip

\noindent
\textBF{Affiliations:} \hspace{0.7mm} None

\bigskip

\smallskip

\noindent
\textbf{Terminator}
\small
\bigskip

\noindent
\textBF{Title:} Simulating Realistic Bokeh Rendering with an Improved Dataset and Robust Network

\smallskip

\noindent
\textBF{Members:} \hspace{0.7mm} Max Zheng \,--\, \footnotesize{1843639867@qq.com}, \small

\hspace{10.5mm} Hulk Wong, Jay Zou

\smallskip

\noindent
\textBF{Affiliations:} \hspace{0.7mm} None

\bigskip

\smallskip

{\small
\bibliographystyle{splncs04}
\bibliography{egbib}
}

\end{document}